%% file: scs20paper.tex
\documentclass{scspaperproc}

\usepackage{url}
\usepackage{graphicx}
\usepackage{flushend}
\usepackage{ragged2e}
\usepackage{booktabs}
\usepackage{tabularx}

\usepackage{latexsym}
\usepackage{graphicx}
\usepackage{mathptmx}

\usepackage{amsmath}
\usepackage{amsfonts}
\usepackage{amssymb}
\usepackage{amsbsy}
\usepackage{amsthm}

\newcolumntype{R}{>{\raggedleft\arraybackslash}X}
\newcolumntype{L}{>{\raggedright\arraybackslash}X}

\usepackage[pdftex,colorlinks=true,urlcolor=blue,citecolor=black,anchorcolor=black,linkcolor=black,bookmarks=false]{hyperref}

\usepackage{hyphenat}
\newtheoremstyle{scsthe}
{8pt}
{8pt}
{\it}
{}
{\bf}
{.}
{.5em}
{}

\theoremstyle{scsthe}

\newtheorem{definition}{Definition}

\newtheorem{prop}{Proposition}

\begin{document}

%
%

\pagestyle{fancyplain}

\thispagestyle{plain}
\firstPageHead{}

\chead{\fancyplain{}{\itshape\small Yacoub \vspace{8pt}}}

\rhead{}
\cfoot{}
\renewcommand{\headrulewidth}{0pt} 

\input{scsprocbib.tex}           

\setlength{\baselineskip}{12.7pt}





\title{VIRTUAL COMMUNICATION STACK: TOWARDS BUILDING INTEGRATED SIMULATOR OF MOBILE AD-HOC NETWORK-BASED INFRASTRUCTURE FOR DISASTER RESPONSE SCENARIOS}

\author{
\\
Aznam Yacoub \\ [12pt]
Polytechnique Montreal, Heterogeneous Embedded System Laboratory, Montreal, QC, Canada \\
Aix Marseille Université, Université de Toulon, CNRS, LIS, Marseille, France \\
aznam.yacoub@polymtl.ca\\
}

\maketitle

\section*{Abstract}

Responses to disastrous events are a challenging problem, because of possible damages on communication infrastructures. For instance, after a natural disaster, infrastructures might be entirely destroyed. Different network paradigms were proposed in the literature in order to deploy adhoc network, and allow dealing with the lack of communications. However, all these solutions focus only on the performance of the network itself, without taking into account the specificities and heterogeneity of the components which use it. This comes from the difficulty to integrate models with different levels of abstraction. Consequently, verification and validation of adhoc protocols cannot guarantee that the different systems will work as expected in operational conditions. However, the DEVS theory provides some mechanisms to allow integration of models with different natures. This paper proposes an integrated simulation architecture based on DEVS which improves the accuracy of ad hoc infrastructure simulators in the case of disaster response scenarios.

\textbf{Keywords:} DEVS, Simulation Tools, Mobile Ad Hoc Networks, Verification and Validation, Disaster Response.

\section{INTRODUCTION}

Mobile Ad Hoc Networks (MANETs) have been proposed in the litterature \shortcite{KIESS2007,Reina2015,Mohammed2019} as a communication technology in the case of emergency and disasters. Indeed, cellular-based infrastructures might become unavailable due to important damages. While MANETs can be quickly deployed without fixed infrastructure, setup or prior requirements, their flexibility is attractive when communications between victims and rescue teams are crucial. However, their implementations face an important challenge: proving that they are enough reliable compared to other approaches \shortcite{KIESS2007}. While Verification and Validation (V\&V) using real experimentations in emergency conditions is utterly impossible, simulation is an important tool in the MANET research community.

\noindent Simulators are an inexpensive manner to evaluate the performance and the accuracy of algorithms and systems without the use of the actual hardware. Also, simulators allow checking the capacity of a network in extreme conditions by varying various parameters in a virtual way and checking different scenarios \shortcite{Manpreet2014}. However, although their use and development increased, the credibility of their results decreased over the time \shortcite{Kurkowski2005,Hogie2006}. Among the problems encountered during the development of MANETs, some are inherent to simulation in general: repeatability, consistency, and accuracy of the models \cite{Sargent2001}. Particularly, simulators generally focus only on some aspects of the network structure itself without taking into account the complexity and the heterogeneity of the systems which rely on this network: autonomous vehicles, unmanned aircraft systems, communication software, etc. 

\begin{figure}[ht]
	\centering
	\includegraphics[width=11cm]{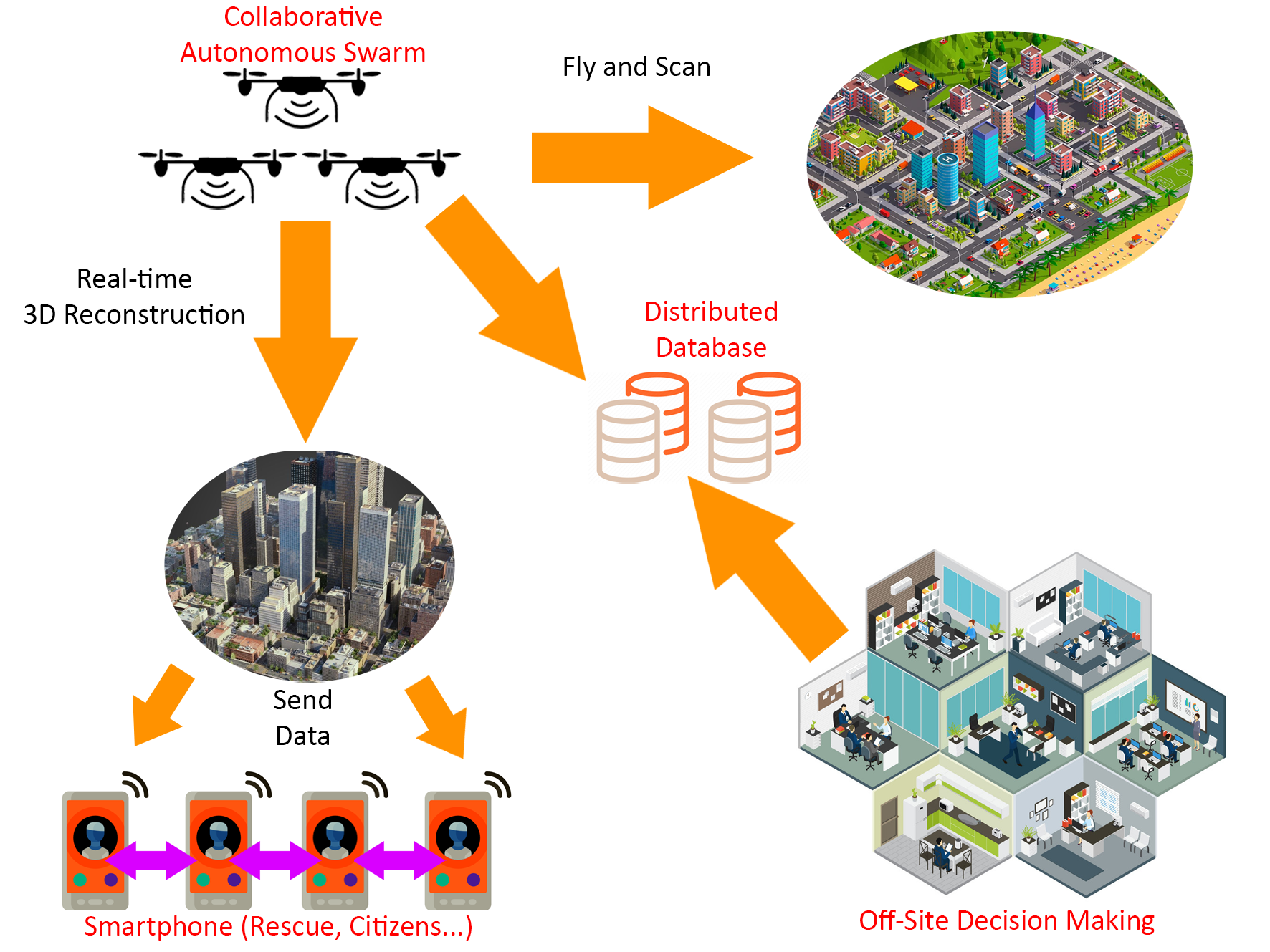}
	\caption{An example of heterogeneous ecosystem communicating using MANET.}
	\label{fig:humanitas}
\end{figure}

\noindent For instance, Figure \ref{fig:humanitas} shows an exemple of real MANET-based ecosystem in an emergency situation. Collaborative drones evolving in a complex environment must communicate without a fixed network infrastructure, send data to different rescue teams with real-time 3D processing software on mobile devices in order to allow professionals to evaluate the situation. Then, these data should also be saved in a database connected to internet in order to allow management teams to take important decisions. Decision support can also be assessed thanks to an Artificial Intelligence-Driven Decision Making Process \shortcite{Phillips2006}. In other words, the verification and the simulation of the entire ecosystem should take into account all the different aspects and natures of all the devices and disasters. This is obviously impossible, but \textit{abstraction} is admitted as a real problem especially in the case of MANET simulation \shortcite{Hogie2006}. Moreover, complex and heterogeneous collaborative systems imply the use of various kinds of models. Therefore, some of these components can be modelled using discrete-event models, continuous models, automata, etc. 

\noindent Various areas addressed the problem of making heterogenenous simulators coexisting and working together in order to improve the accuracy of simulations, and to deal with repeatability and consistency. Essentially techniques like cosimulation \shortcite{Vaubourg2015,Gomes2018} have been proved as good approaches that allow modelers to take into account specificities of different subsystems. However, problems related to repeatability, accuracy and scalability must still be resolved in the case of MANET simulators as stated in the previous cited articles. Especially, the Theory of Modelling and Simulation (TMS), and in particular the Discrete-Event System Specifications (DEVS) formalism \shortcite{Zeigler2000,Zeigler2019}, provides foundations which allow the building of heterogeneous simulators in a hierarchical manner. This kind of simulators allows embedding models with different levels of abstraction, if they respect the DEVS principles. If all the surveys stated in this article show that MANET simulators generally use a discrete-event paradigm, there were few attempts to apply the TMS and DEVS in the case of MANET modelling and simulation.

\noindent In this paper, we propose general guidelines and insights to improve the simulation of MANETs-based infrastructure by using DEVS approaches and results. We show that some results in the DEVS area can benefit to MANET simulation by making easier cosimulations and by making different levels of abstraction coexist inside a unique integrated environment. The first section recalls the existing work concerning MANET simulators and DEVS architecture. In the second section, we introduce our proposed integrated simulation architecture which allows switching between Software-in-the-Loop (SIL) and Hardware-in-the-Loop (HIL) paradigms \shortcite{Murray-Smith2012} by using the DEVS Bus concept \shortcite{Kim1998,Kim2003}. Implementation details and detailed examples are outside the scope of this paper, and are developed in another one.

\section{RELATED WORKS}

Literature about MANET simulation can be splitted in two essentials parts. The first one concerns the common used architectures in the case of network simulation. The second one concerns the actual techniques known in the area of Modelling and Simulation (M\&S), DEVS and simulation theory.

\subsection{MANETs Simulators Architectures}
MANETs simulation can be essentially overviewed by looking at the surveys \shortcite{Manpreet2014,Dorathy2019} which show that simulator architectures have intensively been studied \shortcite{Kurkowski2005,Mallapur2012,Chengetanai2015} without having really evolved for decades. Indeed, as stated by \shortciteN{Kurkowski2005,Andel2006}, this kind of network represents a challenge for simulation community. The complexity is mainly induced by two specific aspects \shortcite{Gunes2007}:

\begin{itemize}
	\item The first one concerns the frequence of the topological modifications. Topology of the network in the case of MANET is related to mobility. However, mobility models generally rely on unrealistic assumptions (randomizations, Manhattan model, etc.). While mobility was understood as having a non-negligible impact on the accuracy of simulations \shortcite{Schindelhauer2003}, mobility models have been specifically studied in separate works \shortcite{Sichitiu2009,Khairnar2011}.
	\item The second one concerns the modelling of physical phenomena. Indeed, MANETs generally rely on wireless communication, which implies radio propagation modelling. As stated in \shortcite{Hogie2006}, study of waves propagation is a complex problem which needs elaborated techniques \shortcite{Schmitz2006}. These techniques make simulations run slower. Therefore, a lot of MANET simulators make strong assumptions and provide simplified models for physical interactions \shortcite{Andel2006}.
\end{itemize}

Testbeds \shortcite{Muchtar2018} can help researchers to overcome these two problems by making real experiments and consider these two models as controlled parameters. However, testbeds remain not scalable and cost expensive.

\noindent Therefore, these two problems are generally abstracted and the tradeoff between reduced accuracy and exe-cution speed is considered as acceptable. MANET researchers therefore develop simulation models which focus on two specific aspects: network performance tests and routing protocols comparison \shortcite{Andel2006}. In the cited article, the authors point that this approach makes wireless models suffer essentially of a lack of accuracy and inconsistencies with extreme divergences between simulators \shortcite{Cavin2002}. If we go further and beyond the existing analysis, we can easily understand the main reasons.

\begin{figure}[ht]
	\centering
	\includegraphics[width=8cm]{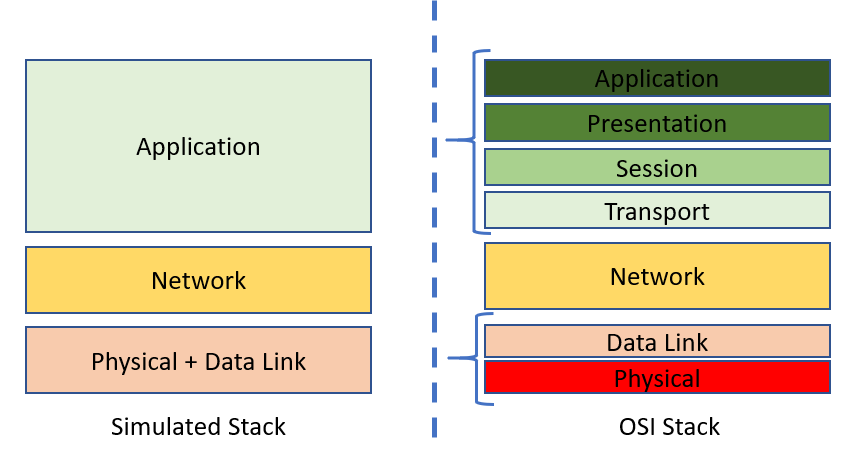}
	\caption{Simulation Stack (left) vs Real Stack (right).}
	\label{fig:stack_comparison}
\end{figure}

\noindent First, \shortciteN{Andel2006} point out the use of unrealistic application traffic during simulation. Indeed, whereas \shortciteN{Hogie2006} states that 
\begin{quotation}
	\textquoteleft Software layers are relatively easy to re-implement within simulators \textquoteright,
\end{quotation}
most of existing MANET simulators commonly use a three-layers protocol stack model (Figure \ref{fig:stack_comparison}) \shortcite{Tuncel2016} which includes:
\begin{itemize}
	\item A physical layer combining the physical layer and the data link layer;
	\item A network layer;
	\item An application layer combining all the layers above the network layer.
\end{itemize}

This architecture could be a good abstraction if the cases of study were representative of the complexity of a software. However, \shortciteN{Andel2006} stated that in most of studies, a constant-bitrate traffic generator is used in the layer 3 of the simulated stack while real software generally more depend on several complex interactions between internal and external components. Especially, traffic generation should take into account a wide range of parameters which depends on usage profile, application specificities, application performance, etc. This statement leads to questionning the assumptions made from the simulation using a \textit{simplified} stack. If we don't question the abstraction of the application layer, we can really wonder what is the impact of this kind of simplification on the accuracy. Especially, in a fully connected environment with heterogeneous components, misuses of protocols can harmfully reduce the performance and can lead to question the robustness of a network. By pushing the reasoning further, interactions between stacks are even more complex in a virtualized environment, in which several stacks can be combined. For instance, in the case of a virtualized operating system, there are at least two communication stacks in the environment. These stacks can be in conflict, resulting a drastically change of the speed in the communication. Therefore, simulating such an environment using one simplified stack could not guarantee the efficiency of the modelled network.

\noindent Second, as we stated before, physical layer is generally modelled as controlled variables or using ideal conditions that are not reflecting the actual implementations. \shortciteN{Takai2001} show the effect of such inaccurate models on the simulation. The experiment consists on evaluating the impact of physical layer settings on Ad Hoc On-Demand Distance Vector (AODV) \shortcite{Perkins1999} and Dynamic Source Routing (DSR) \shortcite{Johnson2001} protocols in simulated environments. These factors signficantly affected the results of the simulation. More important, some settings changed the relative ranking between the protocols. A small variation on the underlying models changed the results of the qualitative comparison analysis between the evaluated protocols. If the result is expected and understandable, it raises an important statement: the level of details is important in the case of MANET simulation \shortcite{Hogie2006} and abstraction or refinement can lead to erroneous outcomes. This opens the way to two fundamental questions: the first one is related to the interpretation of the simulation results, and the second one is related to the V\&V of models.

\noindent Indeed, studies show that this last problem is especially important in the case of MANET simulations. \shortciteN{Kurkowski2005} demonstrate that generally researchers and developers use MANET simulators without prior checking that the used models were validated. Furthermore, when simulation results don't seem to be reliable, they modify the models without new validation steps. As a result, \shortciteN{Andel2006} remind that routing protocols don't properly work in real world in many cases, while they produced good results in simulation. The same observation can also be stated when it comes to verification, at least for Pseudo Random Number Generator (PRNG). If models must be validated against user specifications, algorithm implementations should be also verified against requirements \shortcite{Sargent2011}. 

\noindent Another question is the lack of definition of the good level of abstraction when a model is developed, what \shortciteN{Hogie2006} names the \textit{granularity}, but also from the fact that is utterly impossible to develop enough meaningful scenarios in real-world. Consequently, modellers have no good comparison basis in the case of MANET development, especially in the case of disaster response. Indeed, most of testbed experiments involves less than 50 nodes \shortcite{Hogie2006}, whereas real situations imply hundred or thousand of nodes. Moreover, test and simulation scenarios mainly depend on experts, and imply the question of the coverage, which is out of the scope of this article. V\&V of Simulation Models \shortcite{Sargent2001} are also related to the paradigms and the natures of the different models. Indeed, simulations are often carried from model with different natures, sometimes using discrete-event models, discrete-time models, continuous models, without clearly using a specified formalism. Especially, discrete-event paradigm is well-used for modelling the computational aspects but accurate physical models need continuous approaches. Using the wrong paradigm without motivating it by a serious analysis can lead to increase an undesirable and not necessary heterogeneity of the models. While existing simulators have all a their own purposes \shortcite{Mallapur2012,Manpreet2014} and whereas integrating simulators is a hard challenge, researchers generally focus on one simulator and try to implement models in the paradigm of this simulator \shortcite{Hogie2006}. The outcome of such a practice is the blur of the choice of the good level of abstraction. By extension, it also brings problems while the heterogeneity exists also at the level of implementation, because all of these simulators have their own programming language. In other words, practices in MANET simulation lead to increase the heterogeneity of models over the heterogeneity of the modelled systems, and by extension the number of erroneous outcomes.

\noindent Then, the main problems encountered by MANET simulation studies can be splitted in four categories:
\begin{itemize}
	\item The lack of proper studies when developing the MANET simulators, which leads to inconsistencies; this is more related to methodologies which tend to stick to a simulator instead of trying to develop accurate models for a proper purpose;
	\item The lack of verification and validation in the simulation development process;
	\item The difficult of choosing the right level of abstraction, which leads to oversimplification or overdetailed implementations;
	\item The unrealistic cases of application.
\end{itemize}

MANET simulation community is fully aware of these four problems. If the last one cannot be entirely resolved by a proper development methodology, especially in the case of emergency response, there are clues and insight to resolve the three others. First, \shortciteN{Hogie2006} explicitely show the attempts to develop simulators at the application level: the first one is DIANEmu \shortcite{Klein2003} which provides an environment for simulating applications communicating through a network. However, DIANEmu doesn't simulate the four first layers of the network stack. JANE \shortcite{Frey2004} tries to combine the advantages of simulators, emulators and testbeds by providing a software which is able to work in hybrid mode thanks to a simulation environment and an execution platform. Then, an application can easily swap between the simulated network and the real devices while the communication interface is the same from the point of view of the software. However, the simulation models are themselves defined at a high level of abstraction. Consequently, JANE is not well-suited for a complex heterogeneous environment. However, the results show that the weaknesses of simulation environment introduced in the previous paragraphs can be overcome by introducing emulation aspects. The existence of emulators like JEMu \shortcite{Flynn2001} and MANE \shortcite{Ivanic2009} shows also the importance of the real tests in the evaluation of MANETs.

\noindent A close look at the simulation paradigm used by the well-used MANET simulators \shortcite{Dorathy2019} shows that almost of them implement a discrete approach, in order to reduce the intrinsic complexity of the MANET analytic models. More precisely, some of them like OMNET++ \shortcite{Varga2008} and NS3 \shortcite{Riley2010} use a discrete-event-based architecture without explicitely or fully following the DEVS formalism \shortcite{Zeigler1976}. This is particularly interesting because, as we show in the next section, the TMS \shortcite{Zeigler2019} provides some recommendations which can help to resolve the problems stated previously. Therefore, the next section answers a crucial question: can the DEVS methodology help in more accurate modelling and simulation of MANET and can we provide a methodology to integrate existing MANET simulators into a DEVS-compliant environment ?

\subsection{DEVS Methodology for Modelling, Simulation, Verification and Validation}
\begin{figure}[ht]
	\centering
	\includegraphics[width=3cm]{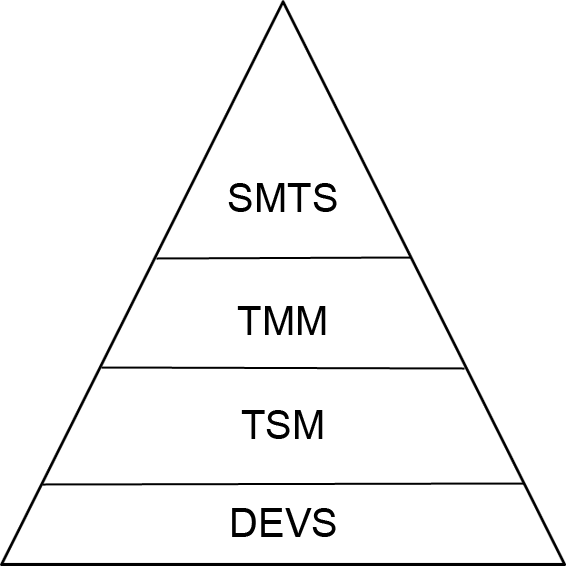}
	\caption{An example of DEVS formalisms hierarchy \protect \cite{Giambiasi2009}.}
	\label{fig:def_hierarchy}
\end{figure}

TMS \shortcite{Zeigler2000,Zeigler2019} gives guidelines for formalizing, modelling and simulating systems in a hierarchical, uniform and universal way. Indeed, the methodology advocates to see any systems as a composition of small black-boxes which take input called observable events, and react according to them. Moreover, each subsystem can also autonomously changes its own state at a specific time \textit{t}, and output a corresponding event. In addition to that, TMS provides a clear separation between conceptual models and computerized (called also simulation) models. More formally, the theory provides a well-defined mathematical specification formalism for structure and behaviour of dynamic systems. DEVS conceptual models are therefore expressed using a clear algebraic structure. Basically, a DEVS model is composed by:
\begin{itemize}
	\item \textbf{a DEVS Atomic model} which is a the most basic unit block. It is a state-machine which describe the behaviour of the component according to received or emitted events;
	\item \textbf{a DEVS Coupled model} which is a composition of DEVS models. Intuitively, it describes the relations and interactions between components.
\end{itemize}
Aside of the algebraic structure of the conceptual model, TMS defines the DEVS Abstract Simulation Model. This model offers the operational interpretation of the DEVS mechanisms. To each DEVS atomic model correspond a DEVS simulator, and to each DEVS coupled model correspond a DEVS coordinator. The DEVS simulator is organized in a tree way, in which the top root coordinator corresponds to the entire model of the system, and each internal node corresponds to a coordinator of a subcomponent. Leafs are the automata simulators which describe the behaviours of the system and subsystems. This architecture allows hierarchical description of models which makes easier the analysis. Furthermore, this clear separation between conceptual and computerized models has many advantages: it allows designers to describe correctly the system under study using the System Modelling Theory and using the good level of abstraction. Indeed, multiple DEVS formalism extensions and subclasses have been developped \shortcite{Giambiasi2006,Giambiasi2009,Hwang2011,Hwang2014} in a hierarchical way. Each extension(resp. subclass) encapsulates(resp. is encapsulated in) another formalism. Consequently, the modelling power of DEVS, meaning the level of abstraction, increases or decreases depending on the chosen formalism (Figure \ref{fig:def_hierarchy}).

\noindent Moreover, the hierarchical construction of these formalisms means that any model expressed in one formalism can be translated to a DEVS model, and each combination of DEVS model is a DEVS model thanks to the closure under coupling property \shortcite{Zeigler1976}. This property enables the interoperability between DEVS-compliant simulators. While it is proved that continuous model can also be encapsulated in DEVS model, it allows also the possibility to mix different paradigms in an heterogeneous simulator.

\noindent Some attempts to use the DEVS formalism in the case of MANET simulation have been done. Especially, \shortciteN{Kim2007,Tuncel2016} proposed fundations for using the DEVS formalism as a basis of MANET simulation. These work show that it is possible to take benefits from the advantages of the DEVS methodology, and that is possible to model scalable, adaptive, reusable, costless and powerful mobile network applications. In the first article, the authors propose to use existing MANET simulators like NS2 for modelling low-level network protocols and components, while a DEVS simulator is used as a controller for high-level behaviours and as a handler of interactions between actors and components. However, this architecture always suffer from the lack of precision of the high-level layer and continue to focus on protocol evaluation. Nevertheless, a main idea raises from these experimentations: it is possible to create an interoperability between a MANET simulator and a DEVS simulator. In the second one, the proposed architecture shows that a full-DEVS simulator can easily simulate MANET protocol as accurate as a network-specific simulator, even using a topology generator.

\subsection{Towards Integrating Non-DEVS Simulator in DEVS-based Architecture}

This last statement is reinforced by the development of a standardized methodology for creating DEVS-based heterogenenous simulation framework \shortcite{Kim1998,Kim2003}. Heterogeneous simulation concerns the use of a collection of simulators developed in different simulation languages and environments, and paradigms, and which work in an interoperable way to achieve a global simulation. A such interoperation needs data exchange and time synchronization between the simulators. While data exchange can be easily resolved through a standard messaging protocol between the simulators, time synchronization is hard because of the possible different natures of the internal models: untimed, continuous, discrete-time or discrete-event. Errors can also come from implementation language which can strongly affects time representation. Parallel and distributed simulations on heterogeneous hardware architecture can also bring errors of approximations.

\begin{figure}[ht]
	\centering
	\includegraphics[width=11cm]{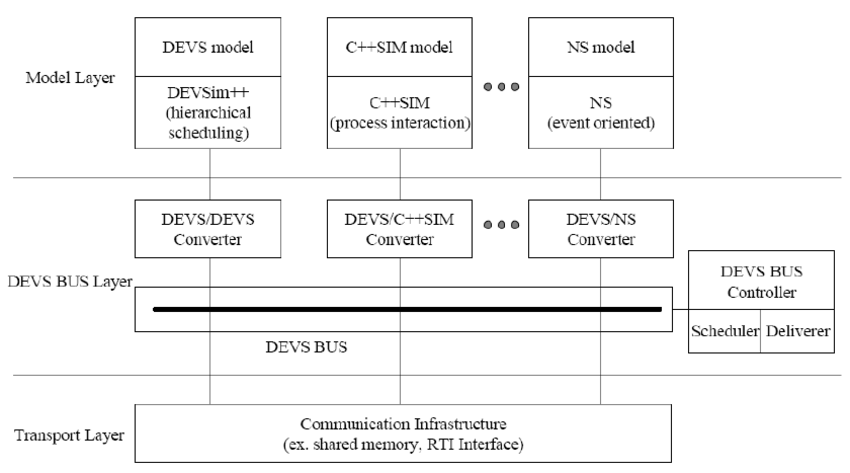}
	\caption{The DEVS Bus developed by \protect \shortciteN{Kim1998}.}
	\label{fig:devsbus}
\end{figure}

\noindent Considering the universality of the DEVS methodology, \shortciteN{Kim1998} developed a DEVS-Bus with the idea that it will provide an unified simulation protocol based on DEVS (Figure \ref{fig:devsbus}). Each simulator is associated to a protocol converter which transforms this simulator into a DEVS-compliant simulator (Definition \ref{def:devscompliant}).

\begin{definition}
	\label{def:devscompliant}
	A DEVS-compliant simulator is a model whose the conceptual model can be defined using a DEVS algebraic formalism, and whose the computerized model follows the DEVS abstract simulator algorithm.
\end{definition}

Therefore, the set of a non-DEVS simulator and its protocol converter implements a DEVS model, which can be coupled with another DEVS model. The entire coupling becomes a DEVS model. Any kind of simulator can then potentially interoperate with any other kind of simulator with a small overhead, if the protocol converter is well-defined and well-implemented. The challenge is then to define a good converter for each simulation integrated protocol. In the case of MANET discrete-event simulator, the task of protocol definition is easy while both DEVS and MANET simulators use the same simulation paradigm. Therefore, the discrete-event structure of the MANET simulator can be coupled to another DEVS simulator, which can be an heterogeneous simulator which uses the DEVS Bus concept. However, the existing approach lacks of proof of correctness and creates a shift between the network topology and the simulation topology. Indeed, the topology of the network communication doesn't necessarily correspond to the structure of the simulation (i.e. the structure of the coupling), since the coupling corresponds to the communication structure between simulators. This can lead to an important overhead while it becomes impossible to evaluate the performance of applications based on the MANET topology under study.

\noindent Taking into account these statements, and the fact that advanced and well-known simulators use a Discrete-Event Architecture, we propose in the next section to integrate them into a DEVS-compliant simulator. We show this approach can be extended to any other MANET simulators while they respect the DEVS principles. Our approach fulfills the following goals adressed by the literature in order to fit to the needs of emergency response simulation:
\begin{itemize}
	\item Using existing MANET models to achieve MANET-based simulation without recreating specific models for our implementation;
	\item Taking into account the specificities of each component of the infrastructure by achieving heterogeneous simulation;
	\item Improving the accuracy of the simulation by executing real software during the simulation and final scenarios, and validating both our simulation model and infrastructure using test cases and use cases;
	\item Allowing swap between Simulation, Emulation, and Execution without redesigning and redeveloping software thanks to a common interface;
	\item Allowing the choice of the good level of abstraction while developing the simulator;
	\item Bringing strong basis for V\&V of MANET simulation models and for V\&V of the resulted infrastructure.
\end{itemize}

\section{INTEGRATED ARCHITECTURE FOR VERIFICATION AND VALIDATION OF MANET-BASED INFRASTRUCTURE}

Our approach will be based especially on the following statements: a DEVS-compliant simulator can be mixed with any DEVS-compliant simulator, and almost everything can be approximated by a DEVS-compliant simulator. First, we define exactly what is a concrete MANET in our case and how it is realized. Then, we show how it can be encapsulated in a DEVS-Bus, and how simulation is finally performed using MANET simulation for the physical part, and the software implementation for the logical part, in order to fulfill our objective to provide an environment which makes possible the verification and the validation of devices and software when they communicate through a MANET.

\subsection{MANET Middleware and Heterogeneous Network Stack}
Basically, the role of a MANET Middleware is similar to the role of a VPN Middleware:
\begin{itemize}
	\item Handle and maintain connections between nodes;
	\item Compute routes between nodes;
	\item Establish and handle communication steps between nodes.
\end{itemize}

\begin{figure}[ht]
	\centering
	\includegraphics[width=9cm]{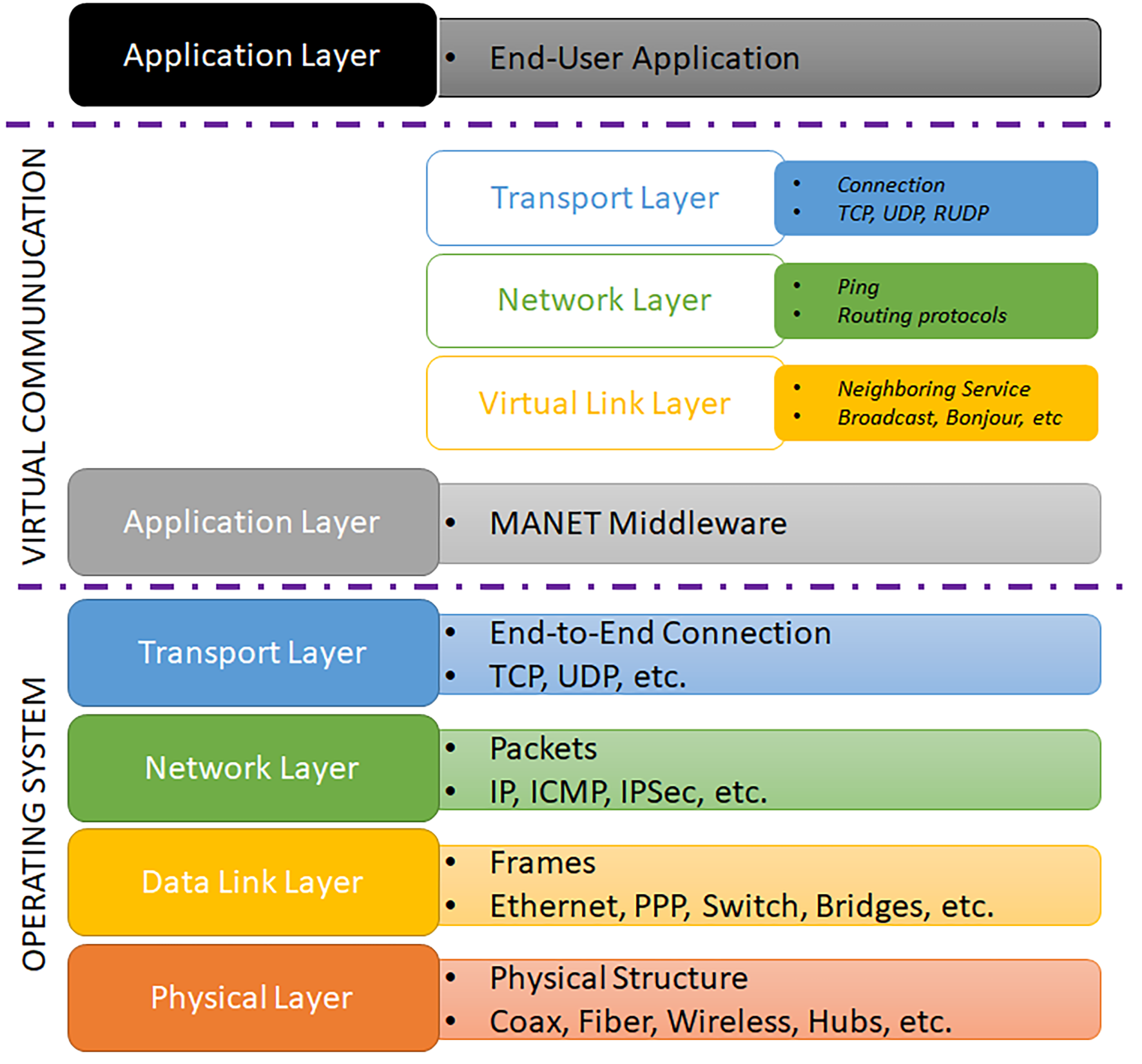}
	\caption{The VCS is between the End-User Application and the OS Stack. The VCS acts as a proxy depending on the execution/simulation mode.}
	\label{fig:heaven_stack}
\end{figure}

Therefore, a MANET middleware can be implemented in multiple ways: virtual stack above the OS stack, bridge between layers of the OS stack, etc. However, in an heterogeneous embedded network, there is as many implementations of the OSI stack as there are devices. We would need as many models of the OS stack as there are operating systems. Consequently, our proposed architecture relies on a Virtual Communication Stack (VCS) (Figure \ref{fig:heaven_stack}) which acts as a proxy and hides specific implementation. The VCS hold the following properties:

\begin{enumerate}
	\item the traffic can goes through the entire stack or can be directly routed to the corresponding OS layer. For the end-user application, data transmission is entirely transparent;
	\item the VCS follows a discrete-event architecture;
	\item the VCS can be embedded in one or several services which are installed on each device which wants to access the network;
	\item each service can be distributed over all the devices (meaning that the implementation of the network can differ according to the platform and the nature of the device);
	\item communication is carried over virtual sockets which acts as normal sockets.
\end{enumerate}

Once the VCS runs on a device, communication are done through the OS stack and through the VCS. Indeed, on the one hand, when a message is received by the OS, it is routed to the VCS which dispatch the message to the corresponding application. On the other hand, when an application sends a message, the message goes through the VCS in order to compute the receiver, before going in the OS stack to be sent through the network. Therefore, our simulation environment has to integrate at least two network stack: one to simulate the OS stack, and one to simulate the virtual stack. Considering all the possible configurations, all the possible implementations on different hardware, and all possible physical phenomena, modelling these two stacks represents a great challenge. Instead of modelling the middleware, we take advantage of the DEVS-Bus architecture.

\begin{prop}
	Given the automata of a middleware application, we can define a DEVS atomic model which is exactly this automata.
\end{prop}
Intuitively, at a high level of abstraction, we can define a conceptual model of a communication middleware as an automata with synchronization mechanisms. The VCS is a discrete-event model by nature (Property 2), with three kind of events: sending, receiving and updating the routing table. At the lowest level of abstraction (the code level), it is a program, i.e. a finite state machine in which each instruction are executed sequentially. Between two executions, the application remains stable. Date of events are decided by the OS scheduler or the CPU clock. In other words, at the lowest level, a middleware is already at least a discrete-time model of our application, and therefore can be see as a discrete-event model.
We can also demonstrate this structure is really a system by proving the legitimacy property \shortcite{Zeigler2000}. However, we can also understand it intuitively with the hierarchy of formalisms.
As a consequence, it is not necessary to transform our middleware into another DEVS model, while writing a DEVS proxy is only needed.

\begin{prop}
	Given the automata of the OS, we can define a DEVS atomic model whis is exactly this automata.
\end{prop}
The explanation is the same as previously. As a consequence, while the virtual stack calls the OS stack, we can see both of them as a monolithic DEVS model.

A first interesting result appears: while the both stack are already a DEVS model, we can see the whole as a virtual DEVS machine. More precisely, if we can redirect the traffic using a TAP/TUN bridge, therefore it is possible to analyze precisely the communication using the real software instead of a model of the client application. This is close to the TAPBridge functionnality proposed by ns3. Otherwise, we can also make abstraction of the OS layer by modelling it and replacing it by a DEVS model. The choice of the level of abstraction can be done transparently according to the desired configuration.



\subsection{Integration of MANET Simulator and Interoperability}

\begin{figure}[ht]
	\centering
	\includegraphics[width=10cm]{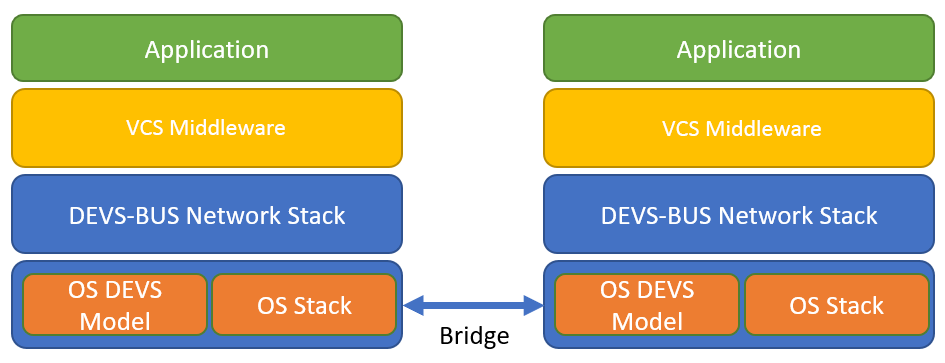}
	\caption{VCS Simulation Environment.}
	\label{fig:firststack}
\end{figure}

The second part of our architecture concerns the integration of existing MANET simulator in order to simulate the physical layer or the OS Stack. For that, we implement two DEVS-Bus (Figure \ref{fig:firststack}):
\begin{enumerate}
	\item The first one translates and synchronizes event from the VCS. In simulation mode, output are filtered according to time. In emulation mode, output are not modified (they are the result of the execution of the algorithm).
	\item The second one translate and synchronizes the input and output of the integrated MANET simulator, in order to make it communicate with the VCS. In emulation mode, this bus send datas directly to the physical media. In simulation, data are sent to the corresponding simulaor.
\end{enumerate}

With this modification, communication are modified according to the environment. Therefore, modellers can choose to test their application in execution, emulation or simulation mode. In the first one, the simulators are disabled. The application runs as in operational conditions. In simulation mode, the DEVS-Bus updates the VCS and consequently the end-user application according to the event of the underlying simulator.

\subsection{Integrated Software-in-the-Loop and Hardware-in-the-Loop Paradigms}
This aspect of our proposed architecture is the use of software-in-the-loop (SIL) and hardware-in-the-loop (HIL) tests. SIL is basically the use of a software instead of a piece of hardware in the test. It is a basic simulation as presented until now or a complete emulation. In contrast, HIL consists on using real piece of hardware in a closed simulation environment. In both of cases, some parts of the architecture can work in \textit{executed time} while others work in \textit{simulated time}. This mainly leads to a synchronization problem: the date of the next event (for instance, $d=10ms$) is long after the time needed to compute it ($t=3ms$). To overcome this problem, we implement a controller in the Testing Environment (SIL/HIL). This controller slows down and bufferize the events and transmit them at the correct date to the hardware. In this way, our simulation environment can embbed real piece of software or real hardware, and work in a mixed mode. This allows us to compare model of our physical part (software or devices) with their real implementation without reimplementing all the architecture.
%
%

From the point the of view of a end-user application, the communication is entirely transparent. Normal application can then be used to test all the protocols of the network, and then compared with the result obtained in real conditions in order to perform triple validation: validation of the simulation model, validation of the MANET-based infrastructure, and validation of the application.

\subsection{Verification and Validation of MANET-based Infrastructure using Integrated Simulation}

Another advantage of such an architecture is that we can decide at any steps of the simulation development if we want to use the final software, or a model of this software. If the software is developed using a verifiable and simulable formalism, we can easily ensure its accuracy in the good context. For instance, it is possible to take advantages of the features of combined V\&V processes. Indeed, DEVS modelling procedure involves classical V\&V processes as defined in \shortcite{Sargent2001,PMBok2004} both for conceptual and simulation model. Moreover, if it was proved a general DEVS model cannot be formally verified \shortcite{Dacharry2005,Dacharry2007}, some subclasses of DEVS models can be translated into formal models \shortcite{Hwang2006,Hwang2009} on which model-checking can be applied. On the another hand, \shortciteN{Yacoub2016} shows that formal models can be combined with DEVS models to take advantages of formal methods and simulation. In this case, a formal model can be translated and augmented with a simulation model. The new obtained model can be then verified and validated using formal verification and formal validation for static properties, and using simulation for checking dynamic behaviours. Integrated in a new software development life cycle, this statement increases the accuracy of the model and of the final software. Indeed, if a final software is built from a verified and validated conceptual and simulation model, it will work as expected designed and tested. Moreover, the capability of using formal methods increases the degree of confidence while the entire statespace can be explored, at least for time-independant properties and invariants. Simulation scenarios can also be easily defined through another External Co-simulation Environment, and integrates more aspects than only the protocol or network parameters.

\section{CONCLUSION}

Disastrous events generally lead to the deployment of reliable network which will be used by heterogeneous systems and software in order to achieve together critical missions. Simulation of MANETs is used to make these networks reliable but actual research focus only on the accuracy of physical and network models, without taking into account the complexity of application, devices and systems. As a result, they lead to erroneous outcomes and the designed network becomes inoperative in real conditions. We propose some insights to build a new simulation architecture based on the DEVS theory in which MANETs simulation can be used jointly with external simulators, SIL and HIL paradigms in order to answer four main problems adressed in the litterature:
\begin{itemize}
	\item the granularity, meaning the choice of the right level of details, through the VCS;
	\item the accuracy implied by the abstraction of the physical model, through the SIL/HIL Testing Environment;
	\item the V\&V of the simulation model during the development process, through the properties of DEVS models;
	\item the use of realistic cases of application when developing MANET-based infrastructure, through the fact that real application can be used in simulation.
\end{itemize} 

In our approach, models and real systems can be swapped in a trasparent co-simulation environment and adapted at each step of the development of the simulator. The MANETs simulator acts as an oracle which regulates the communication between the different systems. Therefore, at the first stage of simulation development, simplified model can be used to tune different parameters and ensure the network model works properly. Then, real software and hardware can be integrated in the simulator to check their real behaviour, reducing the bias introduced by the software and hardware models. However, if this approach allows modularity, constraints induced by the DEVS theory can lead to major overload. Indeed, our proposed architecture implies that the network topology corresponds to the coupling topology. However, Classic DEVS doesn't allow removing or adding node during the simulation. Some workaround can be found, but complex mobility with a lot of topological changes is a problem which must be adressed in a future work.

\section*{Acknowledgments}
I thank Alexy Torres, and Julien Carayol (Polytechnique Montreal) for their feedbacks that improve the manuscript.

\bibliographystyle{scsproc}
\bibliography{scs20paper}

\end{document}

%% file: scsprocbib.tex
\makeatletter
\let\@internalcite\cite
\def\cite{\def\@citeseppen{-1000}%
    \def\@cite##1##2{(##1\if@tempswa , ##2\fi)}%
    \def\citeauthoryear##1##2##3{##1 ##3}\@internalcite}
\def\citeNP{\def\@citeseppen{-1000}%
    \def\@cite##1##2{##1\if@tempswa , ##2\fi}%
    \def\citeauthoryear##1##2##3{##1 ##3}\@internalcite}
\def\citeN{\def\@citeseppen{-1000}%
    \def\@cite##1##2{##1\if@tempswa, ##2)\else{}\fi}%
    \def\citeauthoryear##1##2##3{##1 (##3)}\@citedata}
\def\citeA{\def\@citeseppen{-1000}%
    \def\@cite##1##2{(##1\if@tempswa , ##2\fi)}%
    \def\citeauthoryear##1##2##3{##1}\@internalcite}
\def\citeANP{\def\@citeseppen{-1000}%
    \def\@cite##1##2{##1\if@tempswa , ##2\fi}%
    \def\citeauthoryear##1##2##3{##1}\@internalcite}
\def\shortcite{\def\@citeseppen{-1000}%
    \def\@cite##1##2{(##1\if@tempswa , ##2\fi)}%
    \def\citeauthoryear##1##2##3{##2 ##3}\@internalcite}
\def\shortciteNP{\def\@citeseppen{-1000}%
    \def\@cite##1##2{##1\if@tempswa , ##2\fi}%
    \def\citeauthoryear##1##2##3{##2 ##3}\@internalcite}
\def\shortciteN{\def\@citeseppen{-1000}%
    \def\@cite##1##2{##1\if@tempswa, ##2\else{}\fi}%
    \def\citeauthoryear##1##2##3{##2 (##3)}\@citedata}
\def\shortciteA{\def\@citeseppen{-1000}%
    \def\@cite##1##2{(##1\if@tempswa , ##2\fi)}%
    \def\citeauthoryear##1##2##3{##2}\@internalcite}
\def\shortciteANP{\def\@citeseppen{-1000}%
    \def\@cite##1##2{##1\if@tempswa , ##2\fi}%
    \def\citeauthoryear##1##2##3{##2}\@internalcite}
\def\citeyear{\def\@citeseppen{-1000}%
    \def\@cite##1##2{(##1\if@tempswa , ##2\fi)}%
    \def\citeauthoryear##1##2##3{##3}\@citedata}
\def\citeyearNP{\def\@citeseppen{-1000}%
    \def\@cite##1##2{##1\if@tempswa , ##2\fi}%
    \def\citeauthoryear##1##2##3{##3}\@citedata}
%
%
%
\def\@citedata{%
    \@ifnextchar [{\@tempswatrue\@citedatax}%
                  {\@tempswafalse\@citedatax[]}%
}

\def\@citedatax[#1]#2{%
\if@filesw\immediate\write\@auxout{\string\citation{#2}}\fi%
  \def\@citea{}\@cite{\@for\@citeb:=#2\do%
    {\@citea\def\@citea{, }\@ifundefined
       {b@\@citeb}{{\bf ?}%
       \@warning{Citation `\@citeb' on page \thepage \space undefined}}%
{\csname b@\@citeb\endcsname}}}{#1}}%

%
\def\@citex[#1]#2{%
\if@filesw\immediate\write\@auxout{\string\citation{#2}}\fi%
  \def\@citea{}\@cite{\@for\@citeb:=#2\do%
    {\@citea\def\@citea{, }\@ifundefined
       {b@\@citeb}{{\bf ?}%
       \@warning{Citation `\@citeb' on page \thepage \space undefined}}%
{\csname b@\@citeb\endcsname}}}{#1}}%

%
\def\@biblabel#1{}
\makeatother

\newdimen\bibindent
\bibindent=.25in

\def\thebibliography#1{\section*{\refname}\list
   {}{\settowidth\labelwidth{[#1]}
   \leftmargin \bibindent
   \itemindent -\bibindent
   \listparindent \itemindent
	 \itemsep 4pt
   \parsep 0pt
   \usecounter{enumi}}
   \def\newblock{}
   \sloppy
   \sfcode`\.=1000\relax}